\documentclass[useAMS]{mn2e}

\usepackage{times}
\usepackage{psfig}
\usepackage{graphicx}

\def\spose#1{\hbox to 0pt{#1\hss}}
\newcommand\lsim{\mathrel{\spose{\lower 3pt\hbox{$\mathchar"218$}}
     \raise 2.0pt\hbox{$\mathchar"13C$}}}
\newcommand\gsim{\mathrel{\spose{\lower 3pt\hbox{$\mathchar"218$}}
     \raise 2.0pt\hbox{$\mathchar"13E$}}}

\title[Rapid $\gamma$--ray variability in blazars] 
{Constraining the location of the emitting region in {\it Fermi} 
blazars through rapid gamma--ray variability}

\author[Tavecchio et al.]
{F. Tavecchio\thanks{E--mail: fabrizio.tavecchio@brera.inaf.it}, 
G. Ghisellini, G. Bonnoli, G. Ghirlanda
\\
INAF -- Osservatorio Astronomico di Brera, via E. Bianchi 46, I--23807 Merate, Italy
}

\begin{document}


\pagerange{\pageref{firstpage}--\pageref{lastpage}} \pubyear{2008}

\maketitle

\label{firstpage}

\begin{abstract}
We consider the 1.5 years {\it Fermi}/Large Area Telescope light curves
($E>100$ MeV) of the flat spectrum radio quasars 3C 454.3 and PKS 1510--089, 
which show high activity in this period of time. 
We characterise the {\it duty cycle} of the source by comparing the time 
spent by the sources at different flux levels. 
We consider in detail the light curves covering periods of extreme flux. 
The large number of high--energy photons collected by LAT in these events 
allows us to find evidence of variability on timescales of {\it few hours}. 
We discuss the implications of significant variability on such short 
timescales, that challenge the scenario recently advanced in which 
the bulk of the $\gamma$--ray luminosity is produced in regions of the jet 
at large distances (tens of parsec) from the black hole. 
\end{abstract}

\begin{keywords} galaxies: jets  -- galaxies: individual: 3C 454.3 - PKS 1510--089 -- 
radiation mechanisms: non--thermal -- gamma-rays: observations.
\end{keywords}

\section{Introduction}

Powerful $\gamma$--ray emission is a distinctive feature of 
Flat Spectrum Radio Quasars (FSRQs), radio--loud active galactic nuclei 
with the relativistic jet closely oriented towards the Earth.
Gamma rays with energy above 100 MeV from these sources are widely 
believed to be produced through the inverse Compton (IC) scattering between 
highly relativistic electrons in the jet and ambient photons, either 
the optical--UV photons from the accretion disk (e.g. Dermer \& Schlickeiser 1993) 
or reprocessed by the gas in the broad line region (BLR, e.g. Sikora et al. 1994) 
or the infra--red photons from the dusty torus (e.g. Blazejowski et al. 2000). 

The sources of target photons dominating the IC emission is basically 
determined by the location of the emitting region in the jet 
(e.g. Ghisellini \& Tavecchio 2009, Sikora et al. 2009). 
Nuclear optical--UV seed photons dominate if $\gamma$--rays are 
produced inside the BLR (at distances $<$0.1--1 pc). 
On the contrary, if the emission occurs at larger distances, 
the dominating population of target photons will be that from the torus. 
In turn, the location of the $\gamma$--ray production 
and nature of the target photons determines some of the quantities 
derived from the radiative models, such as electron energies and cooling 
times, jet power, magnetic field intensity. 

\begin{figure*}
\vskip -2.7 true cm
\centerline{ 
\hskip 0.7 truecm
\psfig{file=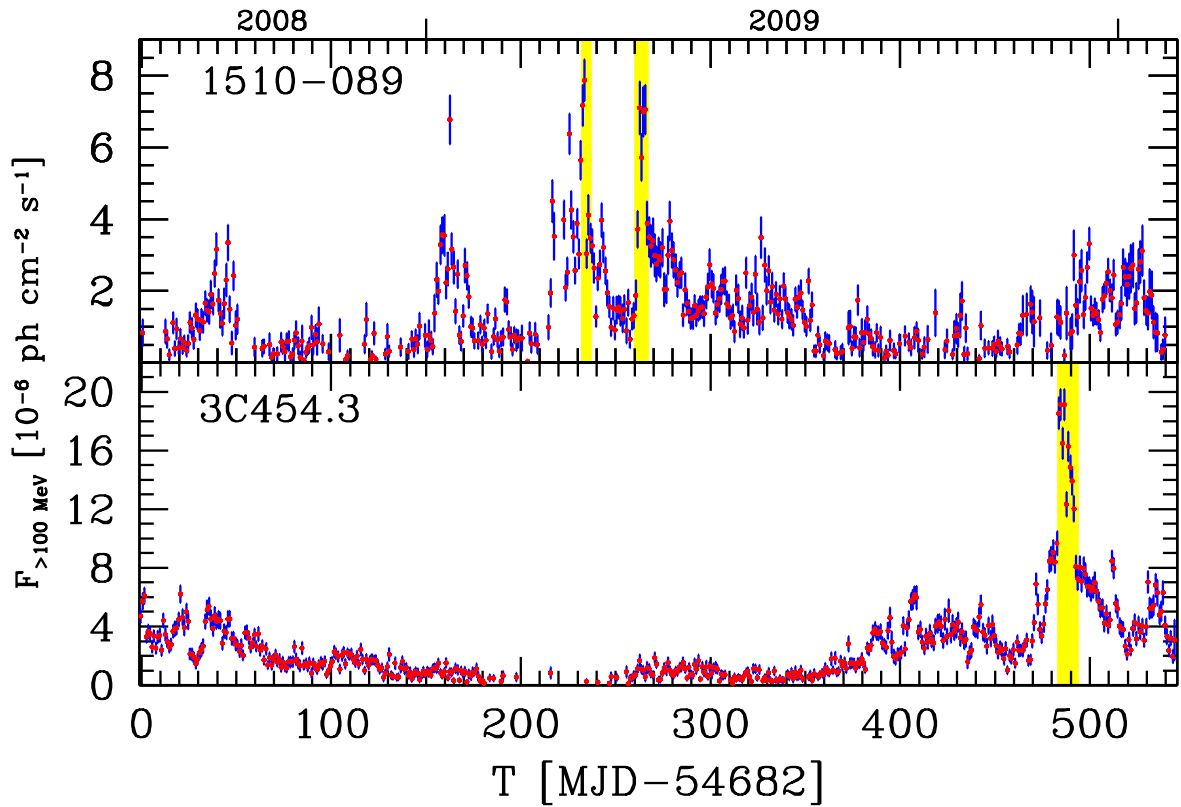,height=17.6cm,width=17.6cm} }
\vskip -3 true cm
\caption{Light curve of PKS 1510--089 (upper panel) and 3C 454.3 
(lower panel) from 2008 August 4 (MJD 54682) to 2010 January 31 
(MJD 55228) in bins of 1 day. 
Vertical yellow stripes show the 
periods of large flux studied in detail in this work.}
\label{1.5anni}
\end{figure*}

The overall spectral energy distributions of FSRQs can be well 
reproduced assuming an emission region located at 300--1000 
Schwarzschild radii from the central black hole (e.g. Ghisellini et al. 2010). 
However, recently, several authors (Sikora et al. 2008; Larionov et al. 2008; 
Marscher et al. 2008, 2010) argued  that the bulk of the emission, especially 
during large outbursts, is produced at larger distances, even at distances of 
the order of 10--20 pc from the central black hole, at the expected location 
of a reconfinement shock (e.g. Sokolov et al. 2004). 
The main arguments advanced to support this conclusion come from 
observations in the radio band coupled with the observed peculiar behaviour 
of the polarisation angle in the optical. 
This led Marscher et al. (2008) to suggest a general scenario in which
blobs (or knots) of material ejected from the central region are 
forced by the magnetic field to follow an helical path, accounting for the observed 
rotation of the polarisation angle in the optical. 
These knots are opaque in the radio band until they reach large distances. 
The transition to transparency 
is marked in VLBI maps by the passage of the compact radio core, after which  
knots become visible and their trajectories can be directly traced. 
The passage of the knots from the core, interpreted as the location of a standing conical 
shock, is marked by the huge flares at all wavelengths, triggered by the 
compression of the plasma in the shock. 
This scenario has important 
consequences for the variability of the emission: since the emission 
region is located at large distances from the central engine, their 
size is probably large, even if a very small jet opening angle ($\theta _{\rm jet}<1$ deg) is assumed. 
Quantitatively, the light crossing time of the source put a lower limit on the variability timescale expressed as:
$t_{\rm var}>\theta _{\rm jet} d (1+z)/c\delta$, where $d$ is the distance of the emission region from the central engine
and $\delta$ the relativistic Doppler factor.  Assuming $d=15$ pc, $\theta _{\rm jet}=0.1$ deg and $\delta=20$ we 
find $t_{\rm var}>1.5 (1+z)$ days.
Therefore, variability at timescales below 1 day, especially at 
$\gamma$--ray energies, would be rather difficult to accommodate in this scheme. 

The Large Area Telescope (LAT) onboard {\it Fermi} (Atwood at al. 2009), with 
its continuous monitoring of the sky, is the ideal instrument to investigate 
the possible existence of rapid (timescale $t_{\rm var}<1$ day) $\gamma$--ray 
variability in FSRQs. 
Indeed, variability on timescale of 12 hours has been 
already reported for the FSRQs PKS 1454--354 ($z=1.424$; Abdo et al. 2009) and PKS 1502+105 
($z=1.83$; Abdo et al. 2010), which underwent major flares in August--September 2008, 
reaching fluxes of the order of a few times $10^{-6}$ ph cm$^{-2}$ s$^{-1}$. 
To further probe variability at short timescales we consider here the 
LAT light curves of two well studied FSRQs, 3C 454.3 ($z=0.859$) and PKS 1510--089 
($z=0.360$), for which the 1--day averaged flux above 100 MeV reached or even 
exceeded in few occasions $10^{-5}$ ph cm$^{-2}$ s$^{-1}$. 
One of these events for PKS 1510--089 has been recently interpreted in the 
framework discussed above (Marscher et al. 2010). 
Such a large flux allows us to investigate variations occurring on 
timescales of the order of  3--6 hours, thus providing strong 
constraints on the theoretical scenarios. 

\section{LAT light curves}

Light curves are derived by analysing the publicly available data with 
the standard Science Tools 9.15.2, including Galactic and isotropic 
backgrounds and the instrument response function P6 V3 DIFFUSE.
For each time bin, we select  the useful events and good-time intervals 
considering a zenith angle $<105$ deg to avoid the Earth albedo and the 
photons within a region of interest with radius 10 degrees centred on 
the position of the source. Then we calculate the livetime, the exposure 
map and the diffuse response. Finally we analyse the data by using an 
unbinned likelihood algorithm (with the task {\tt gtlike}) modelling the 
source spectrum with a power law model, with integral flux (in the 0.1--100 GeV band) 
and photon index as free parameters. 
We plot in the light curves only the time intervals for which the corresponding 
test statistics ($TS$, Mattox at al. 1996; see also Abdo et al. 2009) is larger 
than 10, corresponding to a significance of roughly $\sqrt{TS}\simeq 3\sigma$. 
Due to the intense flux, with this condition only few bins are excluded from 
the light curves.

\begin{figure}
\vskip -1.1 true cm
\centerline{ 
\hskip 0.55 truecm
\psfig{file=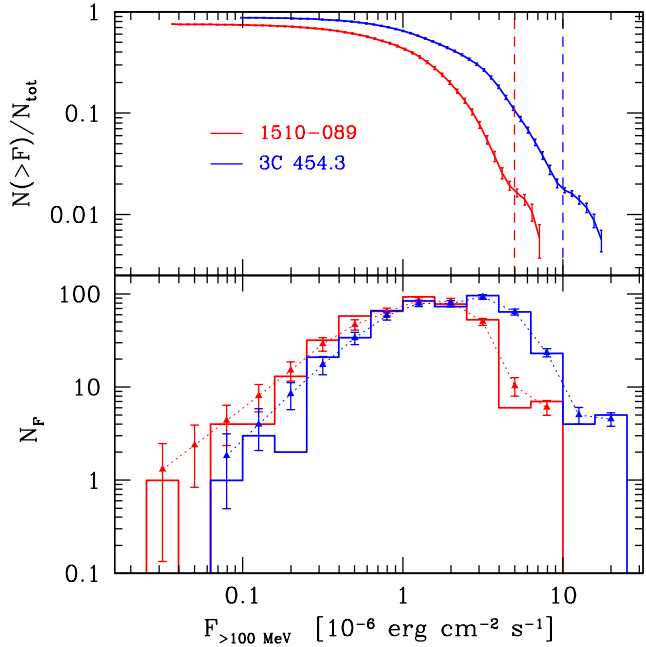,height=10cm,width=9.8cm} }
\caption{
{\it Upper panel}: The number of days (normalised to the total) 
for which the $\gamma$--ray flux above 100 MeV was above a given value $F$, 
as a function of the flux for both 1510--089 and 3C 454.3. 
Error bars report the uncertainties derived with Monte Carlo simulations (see text).
Both sources displays a similar behaviour. 
The duration of the exceptional states studied in this work 
(vertical dashed lines) represents about 1\% of 
the total life of the sources in the considered period. 
{\it Lower panel}: 
the solid histogram shows the flux distribution for both sources. 
Points (triangles) with the associated error bars are obtained by Monte Carlo 
simulations (text for details). 
Both sources display a peaked distribution, well described by an increasing power 
law $N_F\propto F^{1.5}$ up to the the peak and a fast decrease above it. 
The exceptional flares produce a tail at high fluxes (visible also as a bump in 
the integral distribution).
}
\label{duty}
\end{figure}

In Fig. \ref{1.5anni} we report the light curves (in bins of 1 day) of 
PKS 1510--089 and 3C 454.3 covering  about 1.5 years, from August 4 2008 
(when regular LAT observations started) until January 31 2010. 
In both cases the light curves show extended periods of low flux 
level together with shorter period of intense activity. 

\subsection{Flux distributions and duty cycles}

To characterise the variability we derive the distribution 
(differential and integral) of the fluxes of the daily light curve (Fig. \ref{duty}). 
Basically these distributions can be used to characterise the $\gamma$--ray 
{\it duty cycle} of the blazars, usually defined as the fraction of time 
spent by the source in active states (e.g. Vercellone et al. 2004). 
We also take into account 
the (relatively large) errors on the fluxes reporting the average value 
in each bin with the corresponding standard deviation obtained performing 
Monte Carlo simulations of 1000 light curves varying the flux around the 
measured value with a gaussian distribution.  
 
Rather interestingly, for both sources the differential distribution 
(lower panel) shows a well defined peak. Within the uncertainties, below the
peak the distribution appears to be described by a power law with
the same slope ($N_F\propto F^{1.5}$) in both sources.
The peak defines the flux at which the sources spend most of the time. 
Low flux and high flux states are relatively less frequent, though it 
is more frequent for the source to be at lower fluxes than at higher ones.

Above the
peak the distribution displays a rapid decrease, with, 
for both sources, a small excess of events at the highest 
fluxes (corresponding at about 10 times the mean flux), covering almost 
1\% of the total time, clearly related to the huge flares shown by the light curves. 
This can be considered the duty cycle of the sources for these extreme events for the
epoch considered here. 

Some {\it caveats} apply to our results: first of all both sources 
have been chosen because of the extreme activity they displayed in 
the considered past 1.5 years: as such, they cannot probably be considered as 
``typical" sources or in a typical state. 
A better characterisation of 
the flux distribution and duty cycle of the two sources would require to consider
a more extended period of time, including also periods of lower activity. 
Indeed both sources during the EGRET observations (1991--2000) were at a 
lower flux level (see e.g. Nandikotkur et al. 2007, Hartman et al. 1999).  If all flux states 
were considered, the duty cycle of the extreme events would be probably smaller than the 1\% derived above.

A detailed discussion on the derived distributions is beyond the scope 
of this paper. We only note that such a distribution, with a well defined power law from 
low to high fluxes and a rapid decrease at higher flux, is somewhat 
surprising, since log--normal distributions are currently discovered 
analysing X--ray light curves of both accretion powered sources 
(both galactic binaries and AGNs; e.g. Gaskell 2004, Uttley et al. 2005) 
and jetted AGNs (e.g. Giebels \& Degrange 2009 for the case of BL Lac itself). 
A deeper discussion will be presented elsewhere.

\subsection{Rapid variability}

Looking at the daily light curve we selected periods of particular 
high flux (yellow vertical stripes in Fig.1 ), suitable to derive light curves 
in smaller time--bins. For PKS 1510--089 we choose two periods centred 
on MJD 54917 and MJD 54962, in which the source underwent strong 
outbursts ($F>5\times 10^{-6}$ ph cm$^{-2}$ s$^{-1}$ above 100 MeV) 
lasting 4--5 days. 
For 3C 454.3 we consider the recent period of exceptional $\gamma$--ray 
emission at the beginning of December 2009 (MJD 55160-55190). 
The maximum of the daily--averaged flux ($F\sim 2\times 10^{-5}$ ph cm$^{-2}$ s$^{-1}$ 
above 100 MeV) was reached on December 2, 2009. 

\begin{figure}
\vskip -1 true cm
\centerline{ 
\hskip 1.1 truecm
\psfig{file=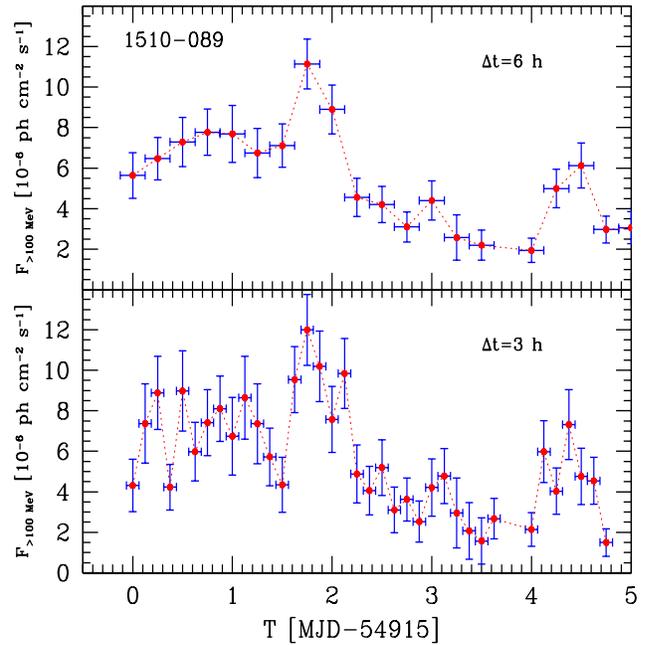,height=10.3cm,width=9.3cm} }
\vskip -0.4 true cm
\caption{Light curve of PKS 1510--089 with bin--size of 6 hours 
(upper panel) and 3 hours (lower panel) starting on MJD 54915 (2009 March 25).
Significant variations with timescales of 6 hours (and marginally also 3 hours) 
are clearly visible.
}
\label{1510-1}
\end{figure}

\begin{figure}
\vskip -0.8 true cm
\centerline{ 
\hskip 1.1 truecm
\psfig{file=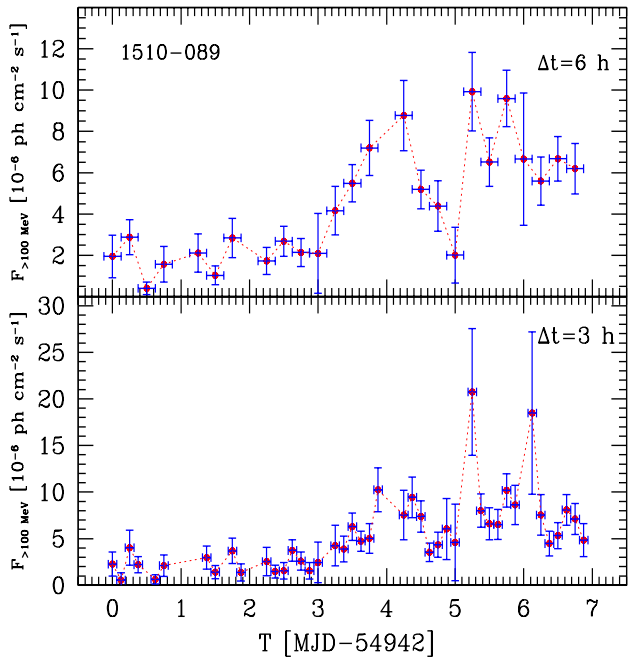,height=10.5cm,width=9.3cm} }
\vskip -0.5 true cm
\caption{As Fig. \ref{1510-1}, starting from MJD 54942 (2009 April 21).}
\label{1510-2}
\end{figure}

\begin{figure}
\vskip -1 true cm
\centerline{ 
\hskip 1.1 truecm
\psfig{file=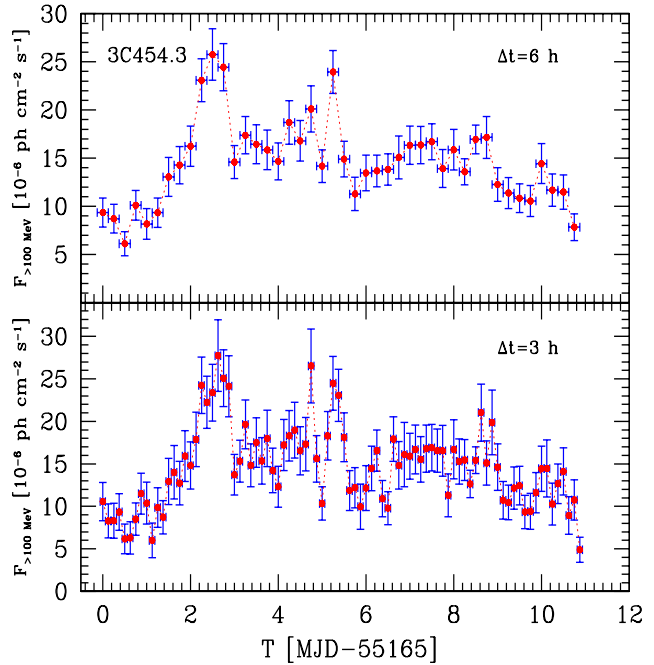,height=10.5cm,width=9.3cm} }
\vskip -0.5 true cm
\caption{As Fig. \ref{1510-1}, for 3C 454.3 starting from MJD 55165 (2009 November 30).
}
\label{454}
\end{figure}

The resulting light curves, with binning of 6 hours 
(upper panels) and 3 hours (lower panels) are shown in Figs. \ref{1510-1}, 
\ref{1510-2}, (PKS 1510--089) and in Fig. \ref{454} (3C 454.3). 
In both sources significant flux variations by a factor of 2 
or more occurring on 6 hours timescale are clearly visible. 
Moreover, there are well defined events (MJD 54916.5 and 
54917.2 for 1510--089, MJD 55167.5 for 3C 454.3) in which even at 3 hours there is evidence 
for variability (both in rising and decaying phases), although the relatively 
large errors prevent a secure conclusion. 

Another feature of these light curve is the approximate symmetry of the peaks, 
(i.e. equal rising and decaying times) suggesting that the relevant timescale  
is the light crossing time of the emission region (e.g. Chiaberge \& Ghisellini 1999). 
However, clearly, the statistics is not sufficient to make definite statements about the flare shapes. For example, the 1510-089 flare on MJD 54946 appears to have a distinctly longer rise than decay time scale.

\section{Discussion}

The evidence of variability on hour time--scale is particular relevant for 
the modelling of the emission of 
the two blazars analysed here in periods of extreme activity. 
Specifically, the condition that the rise and the decay time is less 
than 6 (or even 3) hours puts robust constraints to the size (and consequently 
the location) of the $\gamma$--ray emitting region.
Even considering the most conservative value of 6 hours we can constrain the 
(intrinsic) size of the emitting region to be 
$R<ct_{\rm var} \,\delta /(1+z)=4.8\times 10^{15}(\delta/10)$ cm and 
$3.5\times 10^{15} (\delta/10)$ cm for PKS 1510--089 and 3C 454.3, respectively. 
Even considering extreme values of the Doppler factor, $\delta=50$ 
we obtain values below 0.01 pc. 
This extremely small size is rather difficult to accommodate in 
the picture in which the emission takes place at very large distances 
($\sim$10--20 pc) from the central black hole (e.g. Sikora et al. 2008, 
Marscher et al. 2010), unless the collimation angle of the jet is 
extremely small (Jorstad et al. 2005 derive an extremely small value, 
$\theta _{\rm jet}=0.2$ deg, for both sources; however, using the same technique 
Pushkarev et al. 2009 found larger angles, 2--3 degrees). 
Putting a source with $R$ given by our estimates 
at 20 pc yields $\theta_{\rm jet}\simeq 0.01 \, (\delta/10)$ deg, much 
smaller than the value given by Jorstad et al.
On the other hand, the constraint on the size from the variability 
can be more easily fulfilled if the emission region is located close to 
the black hole, as commonly assumed in the modeling of blazars 
(e.g. Tavecchio et al. 2000, B{\"o}ttcher 2007, Kataoka et al. 2008, Ghisellini et al. 2010; see also 
Bonnoli et al. 2010). 
We conclude that the observation of short variability timescale 
disfavour the ``far dissipation" scenario but can be accommodated in 
the more standard  framework.

Another difference between the two models concerns the different cooling 
timescales of the emitting electrons, determined by the different value 
of the external radiation energy density. 
Specifically, the cooling time of electrons with Lorentz factor $\gamma$ 
as measured in the observer frame is:
\begin{equation}
t^{\rm obs}_{\rm cool} = \frac{3m_e c \,(1+z)}{ 4 \sigma _T \gamma \delta U^{\prime}}
\label{tcool}
\end{equation}
where $U^{\prime}$ is the energy density (radiation plus magnetic) 
as measured in the jet frame. 
The Lorentz factor $\gamma$ can be derived 
by the observed frequency $\nu_{\rm IC}$ of the IC photons: 
$\nu_{\rm IC}\approx \nu_{\rm ext} \gamma ^2 \, \Gamma \delta/(1+z)$, 
obtaining:
\begin{equation}
t^{\rm obs}_{\rm cool} = \frac{3m_ec}{4\sigma_T U^{\prime}} 
\left( \frac{\nu _{\rm ext}}{\nu _{\rm IC}} \right)^{1/2} 
\left( \frac{\Gamma } {\delta }\right)^{1/2} (1+z)^{1/2}
\label{tcool2}
\end{equation}
We assume as typical values for the physical quantities 
$\delta \sim \Gamma =20$ and $\nu _{\rm IC}=2.4\times 10^{22}$ Hz 
(corresponding to an energy of 100 MeV). 
If the emission is produced by the inverse Compton scattering of
external photons within the BLR we have 
$\nu _{\rm ext}=2\times 10^{15}$ Hz and 
$U^{\prime}=3.76\times 10^{-2}\, \Gamma^2$ erg cm$^{-3}$.
Instead, if the seed photons come from the torus we have 
$\nu _{\rm ext}=3\times 10^{13}$ Hz and 
$U^{\prime}=3\times 10^{-4}\, \Gamma^2$ erg cm$^{-3}$ (Ghisellini \& Tavecchio 2008a). 
With these values we obtain observed cooling times of $t_{\rm cool}\simeq 800$ s (BLR)
and $t_{\rm cool}\simeq 12,000$ s (torus).  
Both values are consistent with the observed timescale of variability. 
Therefore we cannot used the observed variability to distinguish between the two cases. 

The same argument can be applied for variability in the X--ray band, in which 
the external IC emission is produced by electrons with much lower energy. 
Assuming $\nu_{\rm IC}=1.2\times 10^{18}$ Hz (i.e. 5 keV), Eq. \ref{tcool2} 
gives cooling times of the order of 31 hours in the case of the BLR scenario, 
and 20 days in the case of the torus. 
Clearly, observations of rapid variability in the X--ray band are difficult to 
explain in the ``far dissipation" scenario.
This discussion implicitly assumes that the observed X--rays come from the 
external IC mechanism; it is possible that in the X--ray band other 
emission mechanism are effective: in particular the Synchrotron Self--Compton 
can contribute, especially in the soft X--ray band (i.e. $\sim$1 keV), where it 
can sometimes dominate the total flux.

Another interesting difference between the two scenarios, that can be 
exploited for further and possibly more conclusive tests, concerns the 
environment external to the region where $\gamma$--ray photons are 
produced and propagate. 
In fact, while in the standard view the $\gamma$--ray 
emission is expected to arise from the inverse Compton emission of soft 
photons of the BLR, at the distances considered in the ``far dissipation" 
model the external radiation fields are dominated by the emission from 
the dusty torus. 
This difference directly translates in a different $\gamma$--ray
spectral shape. 
If $\gamma$--rays are produced within the BLR the combined effects of 
the decrease of the scattering cross--section (Tavecchio \& Ghisellini 2008) 
and of the possible absorption of $\gamma$--rays through pair conversion 
(e.g. Donea \& Protheroe 2003, Liu, Bai \& Ma 2008) 
would result in a spectral cut--off robustly predicted at energies of 
10--20 GeV (e.g. Ghisellini \& Tavecchio 2009). 
In the case of the ``far dissipation" scenario, instead, 
these effects would imprint a cut--off at much larger energies 
($\sim 1$ TeV). 
Therefore, the observation of a spectrum extending up 
to few tens of GeV would  
rule out the standard one--zone model located within the BLR.

One cannot exclude, however,
that more than one component, located at 
different regions from the central black hole, are simultaneously active. 
In this view it is conceivable that inner regions, emitting within 
the BLR, produces the bulk of the fast varying GeV component, while 
more external regions, beyond the BLR can contribute to the emission 
above 100 GeV. 
This could explain the detection of emission above 100 GeV in 3C 279 (Albert et al. 2008) 
and in 1510--089 (Wagner 2010\footnote{HEAD meeting 2010, https://www.confcon.com/head$\_$2010/}). 
If this is would be the case, the overall SED would be complex, with different 
frequency bands dominated by regions at different locations 
(e.g. Ghisellini \& Tavecchio 2009). 
Also the variability would be complex, 
with the possibility to have a rapidly varying GeV emission accompanied by 
a TeV emission varying on much longer timescales. 
We finally note that in 
the current view of FSRQs, the detection of a rapidly varying ($\sim$ hours) 
TeV emission would be quite difficult to explain, since it would require to 
be produced at large distances (to avoid the absorption of photons by the 
optical--UV radiation of the BLR) but in a region not larger than $\sim 10^{16}$ cm. 
Similarly to the case of the ultrafast variability observed in PKS 2155--304 
(Aharonian et al. 2007), this would require to assume the presence of ``needles" 
of emission within the more extended jet 
(e.g. Ghisellini \& Tavecchio 2008b, Giannios et al. 2009).


\section{Conclusion}

We have analysed the LAT light curves of the two FSRQs 3C 454.3 and 1510--089 
from August 2008 to January 2010. 
In this period the two sources displayed several $\gamma$--ray flares with 
fluxes approaching or even exceeding $10^{-5}$ ph cm$^{-2}$ s$^{-1}$. 

We have characterised the variability of the two sources deriving the 
distribution of the fluxes. 
We found that in both cases this distribution 
is well approximated by an increasing power low at low fluxes 
($N_F\propto F^{1.5}$) up to a peak above which the probability rapidly decreases. 
These cases are therefore different from those well described by log--normal 
distributions recently found in both accreting systems and blazars in the 
X--ray band. 

We have derived light curves around the epochs of the largest flares with bins 
of 3 and 6 hours, finding that there are several cases in which the flux varies 
on these timescales, with variations even larger than a factor of 2. 
This implies a correspondingly very compact emitting region
that is difficult to be explained by models that
assume that most of the $\gamma$-ray emission is produced in regions far 
away from the central engine, at distances of tens of parsecs.
The same conclusion, i.e. that the bulk of the emission is produced close to the central engine, was also 
independently reached by Kovalev at al. (2009), using a different approach.
 
\section*{Acknowledgements}
We thank L. Foschini for suggestions on the analysis of the LAT data and comments
on the manuscript. We thank the referee for constructive comments. We thank Y. Kovalev for suggestions.
This work was partly financially supported by a 2007 COFIN-MiUR grant 
and by ASI grant I/088/06/0. This work is based on the publicly available Fermi data obtained 
through the Science Support Center (SSC).

\end{document}